\documentclass[aps,prl,reprint,superscriptaddress]{revtex4-2}
\usepackage{graphicx}
\usepackage{amsmath}
\usepackage{amssymb}

\usepackage{xcolor}
\usepackage{hyperref}

\usepackage{graphicx}
\usepackage{tikz}
\usepackage{pgfplots}
\pgfplotsset{compat=1.18}

\tikzset{
  declare function={
    doublewell(\x,\y) = (\x^2 +\y^2 - 1)^2;
  }
}

\newcommand{\method}{\textsc{HS-REX}}

\begin{document}
\title{Hyperspatial Sampling: Circumventing Free-Energy Barriers\\ via Replica Exchange with Extra Dimensions}
\date{\today}

\author{Henrik Christiansen}
\email{henrik.christiansen@neclab.eu}
\affiliation{NEC Laboratories Europe GmbH, Kurfürsten-Anlage 36, 69115 Heidelberg, Germany}
\author{Matheus Ferraz}
\affiliation{NEC OncoImmunity AS, Forskningsparken, Gaustadalléen 21, 0349 Oslo, Norway}
\author{Takashi Maruyama}
\affiliation{NEC Secure System Platform Research Laboratories, 1753 Shimonumabe, Nakahara-ku, Kawasaki, Kanagawa, Japan}
\author{Francesco Alesiani}
\affiliation{NEC Laboratories Europe GmbH, Kurfürsten-Anlage 36, 69115 Heidelberg, Germany}

\begin{abstract}
Simulating systems with rugged free-energy landscapes remains a central challenge in computational physics and chemistry.
We introduce hyperspatial replica exchange (\method{}), an enhanced sampling method in which the physical system is artificially extended by additional spatial dimensions. In higher dimensions, free-energy barriers can be circumvented through paths that are geometrically inaccessible in the original space.
Restricting the penalty to only solute atoms dramatically reduces the number of replicas required for solvated systems compared to standard temperature replica exchange, a feature especially relevant for biological applications. As proof of concept, we demonstrate the method on a double-well model system and on alanine dipeptide in explicit water as benchmark system. In the latter case, \method{} achieves enhanced conformational sampling of not only the slow backbone dihedral angles, but also both chiral configurations of the molecule, which are sterically inaccessible to standard sampling in three dimensions. This demonstrates enhanced ergodic sampling over conventional temperature replica exchange.
\end{abstract}
\maketitle
The simulation of physical systems in the canonical ensemble at low temperatures is difficult due to many free-energy minima separated by large barriers~\cite{janke2007rugged}.
The goal is to obtain uncorrelated samples from the Boltzmann distribution $p({\bf x})$ at fixed temperature $T$ of Hamiltonian $\mathcal{H}$,
\begin{equation}
p({\bf x}) = \frac{\exp(-\mathcal{H}({\bf x})/k_bT)}{\mathcal{Z}},
\end{equation}
where ${\bf x}$ stands for the spatial degrees of freedom of the system residing in $\mathbb{R}^{N\times d}$, where $d$ is the spatial dimension of the system, $N$ is the number of components (e.g., particles), and  $k_b$ the Boltzmann constant.
Generally, the partition function $\mathcal{Z}$ is intractable.

To sample all relevant free-energy minima, several techniques have been developed that enhance transitions between metastable states~\cite{okamoto2004generalized,mitsutake2012review}.
They can be roughly qualified in two classes:
\emph{i}) Perform importance sampling by simulating an altered Hamiltonian and reweighting the observed configurations to the target distribution, resulting in methods like Umbrella sampling~\cite{torrie1977nonphysical}, Multicanonical~\cite{berg1992multicanonical}, Wang-Landau~\cite{wang2001efficient} or metadynamics~\cite{laio2002escaping}.
\emph{ii}) Simulate a set of systems modeled by different Hamiltonians, chosen in such a way that the (relevant) free-energy barriers vanish in one extreme.
This is done sequentially, such as sequential Monte Carlo (SMC)~\cite{doucet2001introduction} or population annealing (PA)~\cite{hukushima2003population, machta2010population, christiansen2019accelerating} by a combination of sampling and resampling.
Alternatively, it is possible to do this in a simultaneous setting by sampling all individual Hamiltonians and then performing exchanges between them.
The latter approach is known as Parallel Tempering (PT)~\cite{swendsen1986replica,geyer1991markov} or Replica Exchange (REX)~\cite{hukushima1996exchange,sugita1999replica} and is commonly used to enhance sampling both for molecular dynamics and Monte Carlo simulations. 
In PT the systems are simulated at different temperatures, effectively flattening the free-energy landscape and facilitating transitions between metastable states~\cite{earl2005parallel,hansmann1997parallel}.
An extension to this is to scale only part of the Hamiltonian as done in REX~\cite{fukunishi2002hamiltonian,sugita2000multidimensional} or add a guiding potential that vanishes in the target case.

\begin{figure*}[t]
    \centering
    \begin{tikzpicture}
        \node[anchor=south west, inner sep=0] (pdfbg) at (0,0) {
          \includegraphics[width=0.85\textwidth]{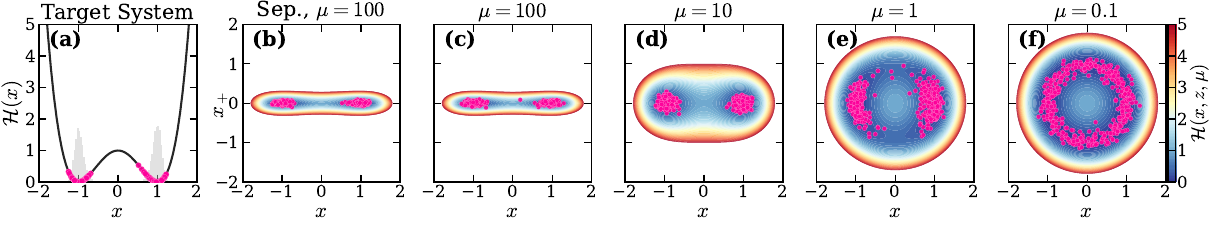}
        };
    
        \begin{scope}[x={(pdfbg.south east)}, y={(pdfbg.north west)}]
          \node[anchor=center, name=plot3d] at (1.1, 0.50) {
            \begin{tikzpicture}[scale=0.7] 
            
              \begin{axis}[
                width=5cm,
                height=5cm,
                view={25}{30},            
                colormap/hot,          
                domain=-1.1:1.1,
                domain y=-1.1:1.1,
                samples=20,
                samples y=20,
                zmin=-0.0, zmax=3,
                axis lines=center,        
                ticks=none,                
                xlabel={$x$},
                ylabel={$x_+$},
                zlabel={$\tilde{\mathcal{H}}(\overline{\bf x})$},
                axis line style={->, opacity=0.7, line width=0.8pt}
              ]
                \addplot3[
                  surf,
                  shader=interp,
                  opacity=0.7             
                ] {doublewell(x,y)};
              \end{axis}
            \end{tikzpicture}
          };

          \node[anchor=north west, font=\bfseries\normalsize,
                fill=white, fill opacity=0.65, text opacity=1,
                inner sep=1.0pt]
              at ([xshift=0.05cm, yshift=0.15cm] plot3d.north west) {(g)};
        \end{scope}
    \end{tikzpicture}

    \caption{(a) Target one-dimensional potential $\mathcal{H}(x)$ (black curve) with the marginal $x$-histogram and trajectory samples (pink) from the separable \method{} replica, showing equal occupancy of both minima.
    (b)--(f) Contours of the two-dimensional hyperspatial potential $\mathcal{H}(\overline{\bf x})$ and trajectory samples for $\mu=100$ (separable), $100$, $10$, $1$, and $0.1$.
    As $\mu$ decreases, the penalty on the extra coordinate $x_+$ weakens, the accessible region in the extended space expands, and samples at small $\mu$ circumnavigate the $x$-barrier via nonzero $x_+$.
    Exchange acceptance is $\sim\!100\%$ between the separable replica and $\mu=100$, and $36$--$45\%$ between neighboring non-separable pairs. The $d=2$ dimensional Hamiltonian is shown in (g).}
    
    \label{fig:double_well_hyperspatial}
\end{figure*}
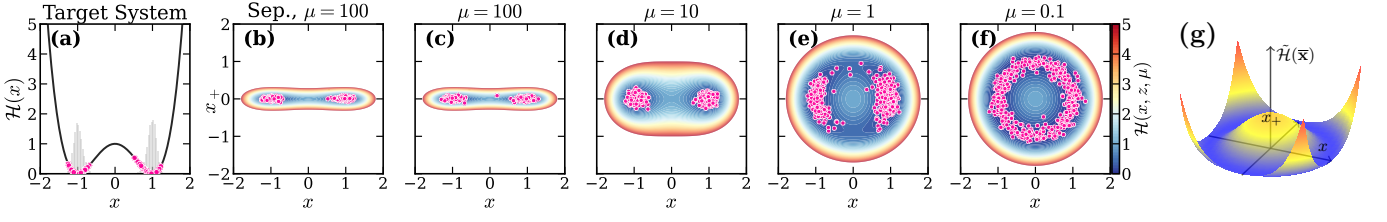
\begin{figure*}[t]
    \centering
    \includegraphics[width=\textwidth]{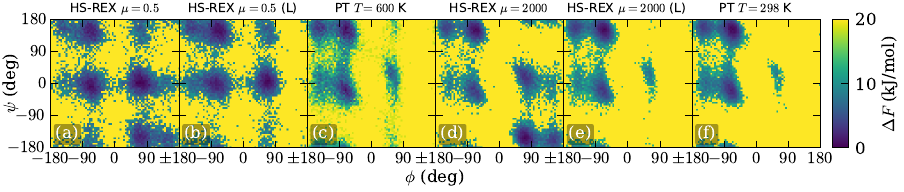}
    \caption{Backbone free-energy surfaces $\Delta F$ for \method{} and PT at the physical and enhanced replicas.
    (a)--(c) show the ability of the methods enhanced replicas to explore the configuration space, (d)--(f) present the sampling under target physical conditions.
    (a) and (d) are results from \method{} at (a) $\mu=0.5$  and (d) the separable system with $\mu=2000$, respectively. 
    Panels (b) and (e) show the same data, but remapped to $L$-chiral configurations by transforming $D$-form frames as $(\phi,\psi)\mapsto(-\phi,-\psi)$.
    Reference PT simulations are show in (c) for $T=600$~K and (f) the target temperature $T=298$~K.
    The data shown uses an identical number of simulation steps. The initial equilibration frames are discarded.}
    \label{fig:ala2_sampling}
\end{figure*}

Despite this arsenal of methods, a fundamental difficulty persists: Any approach confined to three-dimensional space must eventually force the system through the same bottleneck defined by the barrier.
The simulation cost scales exponentially with barrier height, and transitions involving steric clashes, i.e., configurations where atoms must briefly overlap in three-dimensional space, remain geometrically forbidden regardless of the method employed.
\par
In this work, we therefore propose to lift the simulation to a higher dimensional space, i.e., we simulate in a \emph{hyperspatial} setting that is able to \emph{circumvent} barriers.
For this, we define the hyperspatial Hamiltonian $\overline{\mathcal{H}}$ acting on the $D=d+d_+$ dimensional coordinates $\overline{\bf x} \in \mathbb{R}^{N \times D}$ as
\begin{align}
    \overline{\mathcal{H}}(\overline{\bf x}) 
    &= \tilde{\mathcal{H}}(\overline{\bf x} ) + \frac12 \mu \mathcal{L}(\overline{\bf x}_+)\\
    &= \tilde{\mathcal{H}}(\overline{\bf x} ) + \frac12 \mu \sum_{i=1}^{N} \sum_{\delta=1}^{d_+} x_{i,d+\delta}^2
\end{align}
where the higher-dimensional base Hamiltonian $\tilde{\mathcal{H}}$, reduces to the original Hamiltonian $\mathcal{H}$ when all extra dimensional variables are set to zero, and $\overline{\bf x}_+$ corresponds to the $d_+$ extra dimensional coordinates, i.e., $ \bf \bar x = (\bf x, \bf \bar x_+)$. $\mathcal{L}$ is the quadratic energy penalty associated with the extra dimensions.
This definition closely follows the one presented by Pickard in Ref.~\cite{pickard2019hyperspatial} where it was used to facilitate the \emph{optimizations} of structures for atomic clusters, binary compounds, and covalently bonded carbon networks, covering systems with geometric, compositional, and topological frustration.
\par
By simulating $\tilde{\mathcal{H}}$ in the higher-dimensional space, we facilitate barrier crossing.
For example, for classical molecular force-fields, lifting $\mathcal{H}$ to $\tilde{\mathcal{H}}$ can canonically be achieved by measuring distances and angles in the full $D$-dimensional space, which naturally reduces to $d$ dimensions if $\overline{\bf x}_+=0$.
In doing this, atoms can now appear to overlap in the original space and cross large energy barriers geometrically this way.
The role of $\mathcal{L}$ is to penalize the additional dimensions $d_+$, i.e., for $\mu \rightarrow \infty$ one has $\overline{\mathcal{H}}=\mathcal{H}$, while for $\mu \rightarrow 0$ the system is unconfined and lives fully in $D$ dimension.
\par
For the resulting approach of hyperspatial replica exchange (\method{}), we set up a series of systems simulated at the target temperature, but with varying $\mu$, ranging from large penalties to small penalties.
Large $\mu$ in this sense corresponds to the target temperature $T$ in classical PT, and small $\mu$ to the higher temperature $T$ used to cross barriers in PT.
\par
The term $\mathcal{L}$ biases the target distribution. 
In order to sample the correct distribution we can, \emph{a priori}, either perform importance sampling to the final system or simply project the samples into the target system. 
Both of these approaches are unsatisfactory: Either we cannot guaranty large importance weights or that the projected samples are in distribution.
Therefore, we introduce one additional replica at the largest $\mu$ that preserves detailed balance during the exchanges by using the separable Hamiltonian
\begin{equation}
    \mathcal{H}_{\mathrm{sep.}}(\overline{\bf x}) = \mathcal{H}(\bf x) + \frac12 \mu \mathcal{L}(\overline{\bf x}_+),
\end{equation}
which decouples the original system and the penalty term.
In this way, all degrees of freedom are preserved, but the original system is still simulated independently. 
The configuration exchanges can hence be performed without loss of information.
\par
Adjacent replicas use a geometric progression in $\mu$ to obtain approximately uniform exchange acceptance~\cite{kofke2002acceptance,kone2005temperature} (for a derivation, see End Matter).
All replicas are simulated in parallel and exchange configurations are performed according to the standard replica-exchange criterion~\cite{sugita1999replica,fukunishi2002hamiltonian} 
\begin{equation} 
\begin{split}
&p_{ij} = \\
&\min \left( 1, \exp \left[ \frac{\overline{\mathcal{H}}_i (\overline{\bf x}_i) - \overline{\mathcal{H}}_i (\overline{\bf x}_j) + \overline{\mathcal{H}}_j (\overline{\bf x}_j) - \overline{\mathcal{H}}_j (\overline{\bf x}_i)}{k_bT}\right] \right).
\label{eq:rex-acc}
\end{split}
\end{equation}
For replicas differing only in $\mu$ this criterion depends on the penalty term only, and between the separable and non-separable Hamiltonian depends only on the differences in the Hamiltonians; see End Matter for more details.
Because \method{} works via extra dimensions rather than temperature or predefined collective variables, it is orthogonal to many established enhanced sampling schemes and can be straightforwardly combined with them, e.g., with standard PT~\cite{sugita2000multidimensional,earl2005parallel} or metadynamics~\cite{barducci2008well}.
\begin{figure}[t]
    \centering
    \includegraphics[width=\columnwidth]{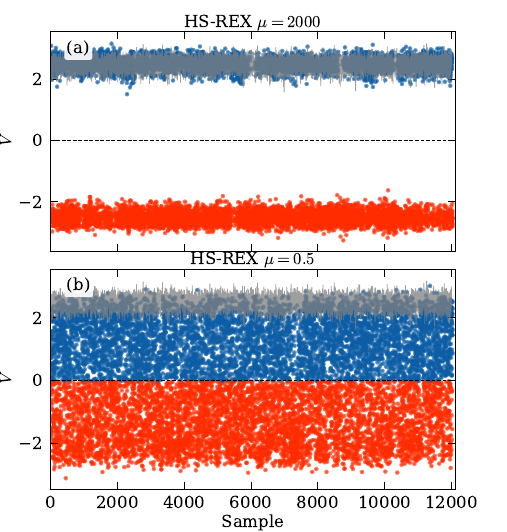}
    \caption{Signed C$\alpha$ chirality volume for alanine dipeptide (common trajectory length; first 100 frames discarded).
    (a) \method{} physical replica ($\mu=2000$); gray: PT $T=298$~K reference.
    (b) \method{} enhanced replica ($\mu=0.5$) with the same reference.}
    \label{fig:ala2_chirality}
\end{figure}
\par
We first demonstrate the approach on a one-dimensional double-well model extended by a single extra coordinate $x_+$,
\begin{equation}
    \mathcal{H}(x) = (x^2 - 1)^2, \quad \overline{\mathcal{H}}(\overline{\bf x}) = \underbrace{(x^2 + \overline{x}_+^2 - 1)^2}_{=\tilde{\mathcal{H}}(\overline{\bf x})} + \frac{1}{2}\mu \underbrace{\overline{x}_+^2}_{=\mathcal L(\overline{x}_+)}.
\end{equation}
The system is simulated at $T=0.1$ where standard Langevin dynamics~\cite{allen2017liquids,leimkuhler2016efficient} with friction coefficient $\gamma=1$ and timestep $\Delta t=0.01$ of the isolated $d=1$ system remains trapped in one minima over extended simulation runs.
Fig.~\ref{fig:double_well_hyperspatial} (a) shows the original Hamilonian $\mathcal{H}(x)$, displaying the minima at $\pm1$.
In (b) the separable system with $\mu=100$ is plotted, and in (c) to (f) we show the extended Hamiltonian $\overline{\mathcal{H}}(\overline{\bf x})$ for different $\mu$.
With decreasing $\mu$ paths open around the barrier in the extra dimension $x_+$ that allow for transition paths with no or small energetic cost.
The pink dots show samples from an \method{} simulation of this ensemble, showing that both minima are sampled with the same frequency in the target system.
This is also reconfirmed in (a), where the equal-height bi-modal histogram is plotted based on the separable Hamiltonian $\mathcal{H}_{\mathrm{sep.}}(x)$.
\par
The choice of the extreme $\mu$ values follows a simple rational: $\mu_{\mathrm{max}}$, used between the separable and non-separable system, needs to be chosen in such a way that a target acceptance rate is achieved between only those two replicas.
The minimal value $\mu_{\mathrm{min}}$, in contrast, is chosen in such a way that all (relevant) energy barriers are crossed.
\par
In addition, we benchmark \method{} with alanine dipeptide (Ace-Ala-Nme) in explicit water.
For this we have extended the differentiable molecular simulator (DIMOS)~\cite{christiansen2025fast} to support higher dimensional classical force fields and implemented standard REX and \method{}~\footnote{Code available at \href{https://github.com/nec-research/DIMOS}{github.com/nec-research/DIMOS}}.
DIMOS is a fully-differentiable \text{pyTorch} based GPU-accelerated molecular simulation package, designed to facilitate rapid prototyping.
To accelerate the simulation, we have also extended DIMOS to support batching of simulations into a \textit{super}-system; more details in the End Matter.
\par
To model interactions, we use the classical force field ff14SB~\cite{maier2015ff14sb} and extend it to hyperspatial coordinates by measuring distances and angles in the full $D$-dimensional space.
Integration is performed using Langevin dynamics and periodic boundary conditions~\cite{allen2017liquids,tuckerman2010statmech}.
The harmonic extra-dimensional penalty is a fast subsystem that is not resolved under a usual timestep of $\Delta t = 0.5$~fs; to tackle this, we integrate partly analytically within a BAOAB--RESPA Langevin scheme~\cite{tuckerman1992reversible,leimkuhler2016efficient}.
For more details on this and the simulation setup, see End Matter.
\par
Simulations of systems with many degrees of freedom traditionally suffer from small inter-replica acceptance rates or the need for many replicas due to very narrow distributions~\cite{kofke2002acceptance,nymeyer2008efficiency}.
Since we, for biological systems, are primarily interested in sampling all free-energy minima of the solute and not the solvent, we can utilize a natural approach \method{} facilitates: 
We penalize the extra dimensions only for solute atoms, analogous in spirit to replica exchange solute tempering (REST)~\cite{liu2005replica,wang2011replica}, but by confining dimensionality rather than scaling interactions.
This has the clear benefit of significantly reducing the number of replicas required for solvated systems compared to standard PT.
Compared to REST which achieves solute-focused enhancement by rescaling interaction strengths, effectively simulating a physically distorted Hamiltonian in enhanced replicas, \method{} instead leaves all interactions intact and opens a geometric path through extra dimensions.
Hence, the target ensemble is never deformed as strongly as in REST.
\par
Hence, we are able to use only 16 replicas with $\mu$ that geometrically span from $\mu_\mathrm{max}=2000$ to $\mu_\mathrm{min}=0.5$.
We compare against PT~\cite{earl2005parallel,hansmann1997parallel} with 16 replicas, using the standard temperature range $T=298$--$600$~K.
All observables are evaluated at an interval of $1$~ps of simulation time per replica after discarding an initial equilibration window corresponding to $500$~ps of simulation time per replica.
\par
Fig.~\ref{fig:ala2_sampling} shows the free-energy surfaces $\Delta F(\phi,\psi)$~\cite{chipot2007freeenergy,maier2015ff14sb} of the backbone at target physical conditions and for the most enhanced replicas for both methods.
The physical replica for \method{} shown in (d) visits both the C$_7^\mathrm{eq}$ basin ($\phi\approx-90^{\circ}$, $\psi\approx150^{\circ}$) and the $\alpha$R basin ($\phi\approx-60^{\circ}$, $\psi\approx-45^{\circ}$), in agreement with the basin locations reported in the literature for alanine dipeptide~\cite{laio2002escaping}.
However, perhaps surprisingly, this plot does not match the reference PT simulation at $298$~K presented in (f).
This is due to a switch in chirality between 
$L$ (left) and $D$ (right) form of alanine dipeptide not resolved by the PT simulation~\cite{periole2007convergence,maier2015ff14sb}.
The chirality switch is directly visible in the signed tetrahedral volume at C$\alpha$,
\begin{equation}
    V = (\mathbf{x}_N - \mathbf{x}_{C\alpha}) \cdot \bigl[(\mathbf{x}_C - \mathbf{x}_{C\alpha}) \times (\mathbf{x}_{C\beta} - \mathbf{x}_{C\alpha})\bigr].
\end{equation}
This order parameter changes sign upon $L/D$ inversion ($V>0$ for the $L$-form, $V<0$ for the $D$-form).
We plot $V$ in Fig.~\ref{fig:ala2_chirality} for \method{} at (a) $\mu=2000$ and (b) $\mu=0.5$.
The gray trajectory shows $V$ for the reference PT simulation.
At $\mu=0.5$, $V$ is sampled nearly uniformly, demonstrating the complete vanish of the barrier, whereas the PT simulation remains clearly in the $L$ form.
\par
Under a classical, non-reactive force field such as \mbox{AMBER~ff14SB}~\cite{maier2015ff14sb}, bonds are not broken or formed during this transition, as the molecular topology is fixed for the entire simulation. 
However, since the force field's bonded terms do not include any energy term that distinguishes the L- and D-form geometries, the two forms are exact energetic and entropic degenerates. 
A correctly ergodic sampler must therefore visit both forms with equal probability.
\par
The barrier separating the two forms in three-dimensional space arises from steric strain rather than from covalent bond breaking. 
Transitions between the L- and D-form forces the molecule through intermediate configurations where non-bonded atoms come into close contact. 
This steric cost renders the transition prohibitively slow for any method restricted to three spatial dimensions, at any accessible temperature. 
The extra spatial dimension in \method{} allows the molecule to interpolate between the two degenerate geometries along a path that avoids these strained configurations. 
This is the mechanism by which the transition becomes accessible at small $\mu$.
It does not correspond, however, to a stereochemical isomerization in the chemical sense, which would require breaking and reforming a covalent bond at C$\alpha$, an entirely different process governed by a much higher energy barrier that is absent from this non-reactive potential.
\par
We reconfirm this by plotting the remapped free-energy surface in Fig.~\ref{fig:ala2_sampling}(e) which shows that both basins are sampled without artificial splitting across enantiomeric halves of dihedral space.
The plot agrees largely with the reference PT simulation at $298$~K shown in (f), with slightly improved sampling in (e) on this timescale, illustrating the backbone isomerization barrier that \method{} overcomes at equal computational cost.
At the enhanced replicas shown in (a) and (c), the two methods behave qualitatively differently.
The corresponding $L$-remapped map of panel (b) shows the same basins merged into a single coherent free-energy landscape, demonstrating that rare $D$-form excursions at the physical replica originate from genuine conformational transitions in extended space rather than from inadequate sampling of a fixed enantiomer.

\begin{figure}[t]
    \centering
    \includegraphics[width=\columnwidth]{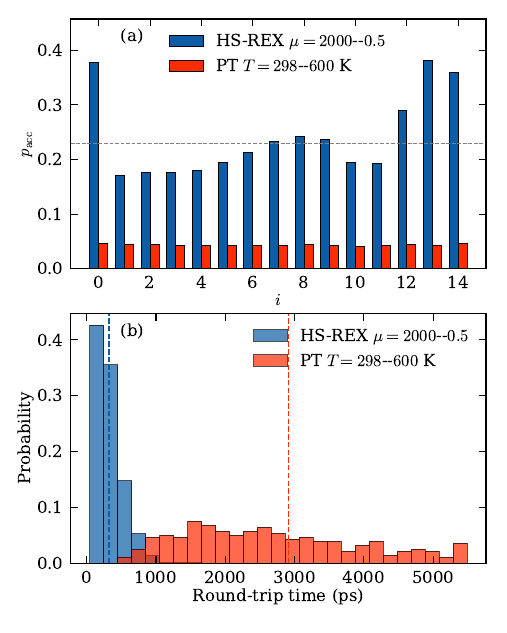}
    \caption{Replica-exchange diagnostics for alanine dipeptide (common trajectory length).
    (a) Per-pair acceptance rates for \method{} ($\mu=2000$--$0.5$, blue) and PT ($T=298$--$600$~K, red); dashed line: optimal $23\%$~\cite{kone2005temperature}.
    (b) Round-trip time distributions (dashed lines: means).}
    \label{fig:ala2_rex_exchange}
\end{figure}
Fig.~\ref{fig:ala2_rex_exchange}(a) quantifies the replica-exchange efficiency.
Because the solute-only penalty leaves solvent atoms in the physical subspace, the number of extra-dimensional degrees of freedom entering the exchange criterion is limited to the solute ($N_\text{dof}=N_\text{solute}$ with one extra dimension per atom), which keeps acceptance computationally tractable in explicit solvent.
\method{} maintains acceptance rates of $17$--$38\%$ across all adjacent pairs (mean $\approx 24\%$), close to the uniform-acceptance optimum of $23\%$~\cite{kone2005temperature} predicted for a geometric $\mu$-ladder (End Matter).
The per-pair rates vary smoothly along the ladder without the catastrophic drop at either end that often plagues PT in explicit solvent~\cite{nadler2008optimized,periole2007convergence}.
By contrast, PT acceptance remains uniformly low at $4$--$5\%$ on every pair (mean $\approx 4\%$): With 16 replicas spanning $298$--$600$~K, adjacent temperature states overlap do not overlap sufficiently for efficient exchange~\cite{kofke2002acceptance,nadler2008optimized}.
\par
The round-trip time distributions in Fig.~\ref{fig:ala2_rex_exchange}(b) reconfirm these observations based on the mixing times for the full replica ladder~\cite{rosta2009error}.
A round trip~\cite{rosta2009error,lingenheil2009efficiency} is defined to start at the physical replica, visiting the most enhanced state, and returning.
It measures how rapidly the ensemble explores the extended state space and delivers decorrelated configurations to the target replica.
\method{} achieves a sharply peaked distribution with mean $324$~ps (median $278$~ps), reflecting fast diffusion along the $\mu$ ladder enabled by near-optimal acceptance.
PT yields a much broader distribution with mean $2625$~ps (median $2312$~ps) and a long tail extending beyond $5$~ns.
This corresponds to an $\approx 8$ times reduction in round-trip times of \method{} compared to PT.
Although the $600$~K replica decorrelates conformations well on short timescales (ignoring the chirality), the replica ladder is rarely transversed; a trajectory initialized at $298$~K therefore requires many rare successful swaps, producing the heavy tail in panel (b).
For larger acceptance rates in PT and smaller round-trip times, more replicas would be needed~\cite{nadler2008optimized}.
\par
In summary, \method{} circumvents configurational barriers by geometric paths in extended spatial dimensions and, with a geometric $\mu$-ladder, provides near-uniform replica-exchange acceptance rates and substantially shorter round-trip times than PT at equal replica count.
On alanine dipeptide in explicit water, this translates into improved sampling of backbone isomerization and rare chirality inversion, transitions that remain inaccessible to standard PT~\cite{earl2005parallel} on the same timescale.
Restricting the penalty to solute atoms makes the approach practical for solvated biomolecular systems by limiting the extra-dimensional degrees of freedom that govern exchange, and generalization to other systems with rugged free-energy landscapes~\cite{janke2007rugged} is a natural next step.
\par
In particular would be interesting to investigate the interplay of \method{} with other approaches, especially with adaptive bias-based methods such as metadynamics~\cite{barducci2008well,barducci2011metadynamics}.
Another direction worth exploring is the influence of the number of extra dimensions $d_+$ on the sampling performance and the number of replicas needed.

\begin{acknowledgments}
We thank Ke Shi and Fabio M\"uller for helpful discussions.
The authors acknowledge the use of Cursor and associated large language models for assistance in code generation and writing of this work.
\end{acknowledgments}

\bibliography{bib}

\appendix

\section{Replica Exchange Criteria}
\label{app:exchange}

The probability of accepting an exchange between replicas $i$ and $j$ follows from the usual replica-exchange criterion~\cite{sugita1999replica,fukunishi2002hamiltonian} presented in Eq.~\ref{eq:rex-acc}.
For exchanges between non-separable replicas with different $\mu$ values, the extended base Hamiltonian $\tilde{\mathcal{H}}$ cancels exactly, so that acceptance depends only on the extra-dimensional penalty energy,
\begin{equation}
\begin{split}
&p_{ij} =\\
&\min \left( 1 , \exp \left[ \frac{(\mathcal{L}(\overline{\bf x}_{+,i}) - \mathcal{L}(\overline{\bf x}_{+,j})) (\mu_i - \mu_j)}{2k_bT}\right]\right).
\label{eq:acceptance_criterion}
\end{split}
\end{equation}
Using similar arguments, the exchange criterion between separable and non-separable replicas at the same $\mu$ reads
\begin{equation}
\begin{split}
&p^\text{sep}_{ij} = \\&\min \left( 1 , \exp \left[ \frac{\mathcal{H}({\bf x}_i) - \mathcal{H}({\bf x}_j) + \mathcal{H}_{\mathrm{sep.}}(\overline{\bf x}_j) - \mathcal{H}_{\mathrm{sep.}}(\overline{\bf x}_i)}{k_bT} \right]\right).
\label{eq:acceptance_criterion_separable}
\end{split}
\end{equation}

\section{Molecular dynamics: Alanine-Dipeptide}
\label{app:md}

All biomolecular \method{} simulations were carried out in DIMOS~\cite{christiansen2025fast}, a GPU-accelerated PyTorch molecular dynamics library we extended for hyperspatial replica exchange.
Initial coordinates for the alanine dipeptide (Ace-Ala-Nme) system were generated using the \texttt{tleap} program from AmberTools~\cite{case2023ambertools}. The solute interactions were described by the ff14SB force field~\cite{maier2015ff14sb}, while the solvent was modeled using the explicit TIP3P water model~\cite{jorgensen1983comparison}. The peptide was placed in a cubic simulation box with a minimum clearance of $10$~\AA{} between the solute and the box boundaries.  Prior to the production runs, the system preparation, including a 50,000-step steepest descent geometry optimization, NVT heating to $T=298$~K for 100~ps, and subsequent 2~ns NPT equilibration, was performed using the GROMACS~2022.5 package~\cite{abraham2015gromacs}. The final coordinates from this equilibrated $d=3$ ensemble were then used as the input configuration for the DIMOS production phase. During production, non-bonded interactions were calculated using a $9.5$~\AA{} cutoff with an $8.0$~\AA{} switching distance, alongside reaction-field electrostatics~\cite{tironi1995reactionfield}. Equations of motion were integrated at $T=298$~K with an outer timestep of $\Delta t=0.5$~fs, regulated by a Langevin thermostat ($\gamma=1$~ps$^{-1}$).
\par
Each atom carries coordinates $\bar{\mathbf{x}}_i\in\mathbb{R}^{D}$ with $D=d+d_+$ and $d_+=1$.
Positions are initialized by padding the equilibrated three-dimensional coordinates with displacements drawn from $\mathcal{N}(0,0.1^2)$~\AA~for $\overline{\bf x}_+$.
Bond lengths, harmonic angles, Lennard-Jones, and electrostatic interactions use the Euclidean distance in $\mathbb{R}^D$; backbone dihedrals are computed from the three-dimensional projection.
Neighbor lists are built from a three-dimensional cell list which constitutes a guaranteed superset because $r_D\ge r_{d}$. 
Replica exchanges are attempted between adjacent replica pairs on an random even/odd schedule after every $100$ MD steps.
\section{Parallelization of Molecular Dynamics}
In conventional replica-exchange workflows, each thermodynamic state is often advanced by an independent process (e.g., one MPI rank per replica or one GPU).
For modest biomolecular systems this leaves modern GPUs underutilized, because a single replica cannot occupy the full device.
An alternative strategy is to \emph{fuse} replicas into one \textit{super}-system: coordinates of $N_r$ copies are concatenated, interaction lists are made block-diagonal so that replica $i$ interacts only with itself, and a single force evaluation advances all replicas in parallel.
This amortizes kernel-launch overhead and improves occupancy without inter-replica communication during the MD phase.
DIMOS implements this pattern for the $15$ non-separable replicas that share identical topology, temperature, and integrator settings and differ only in $\mu$.
Their positions and velocities are stored as tensors of shape $(N_r N_\text{atoms}, D)$; bonded and nonbonded edge lists are offset and concatenated, and nonbonded pair lists are restricted to block-diagonal upper-triangle indices so cross-replica pairs never enter the cutoff sum.
Per-replica penalty strengths are encoded as atom-wise $\mu$ tensors in a batched $\mathcal{L}$ term.
One integrator step therefore updates all non-separable replicas simultaneously on the GPU.
\section{Solute Only Penalty}
To limit the cost of explicit-solvent exchange, the harmonic penalty is applied only to solute atoms.
Solvent atoms therefore remain strongly confined to the physical subspace while the peptide explores the extended coordinate at small $\mu$, reducing $N_\text{dof}$ entering the acceptance criterion to $N_\text{solute}\,d_+$.
Unlike REST~\cite{liu2005replica,wang2011replica}, which scales interaction strengths, this variant changes the effective dimensionality explored by the solute.
\par
Naturally, for other scenarios such as, for example, estimating binding affinities of small drug-like molecules, it is possible to choose other partitions; allowing the small molecule to explore the full higher-dimensional configuration, while keeping the protein highly constrained to the original one.

\section{Multiple-time-scale Integration}
The harmonic confinement $\mathcal{L}$ of the extra coordinates defines a fast subsystem that is a canonical target for reference-system propagator methods.
Refs.~\cite{tuckerman1990condensed,tuckerman1992reversible} introduced reversible reference-system propagator algorithms (r-RESPA), in which a fast reference is integrated with inner sub-steps, and reversible numerical-analytical propagator algorithms (r-NAPA), in which the reference is integrated exactly when it is harmonic.
Our implementation follows the r-NAPA pattern: The slow classical force-field terms are integrated with the outer timestep, while the fast penalty is propagated exactly.
For each extra coordinate $x_+$ with frequency $\omega=\sqrt{\mu/m}$, the exact map over an interval $\Delta t$ is the rotation~\cite{stuart1996mdmts}
\begin{align}
x_+' &= \cos(\omega\Delta t)\, x_+ + \frac{\sin(\omega\Delta t)}{\omega}\, v_+ \label{eq:harmonic-map-z}\\
v_+' &= -\omega\sin(\omega\Delta t)\, x_+ + \cos(\omega\Delta t)\, v_+. \label{eq:harmonic-map-vz}
\end{align}
\par
The Langevin thermostat uses a BAOAB operator split~\cite{leimkuhler2016efficient}:
Slow force kicks (B), fast reference steps (A), and an Ornstein--Uhlenbeck noise and friction step (O).
Ref.~\cite{leimkuhler2016efficient} derived this for the BAOAB operator split for Langevin dynamics combined with RESPA-style solvent--solute splitting.
In our case, the fast and slow partition is between the harmonic penalty and the remaining force field rather than between solvent and solute.

\section{Penalty Parameter $\mu$ Schedule}
\label{app:mu-schedule}
Under harmonic penalty scaling, equal spacing in $\log(\langle x_{i,\delta}^2\rangle^{1/2})$ corresponds to a geometric progression in $\mu$. Motivated by this separable harmonic approximation, we choose
\begin{equation}
\label{eq:mu-geom}
    \mu_i = \mu_\text{max} \left( \frac{\mu_\text{min}}{\mu_\text{max}} \right)^{(i-1)/N}, 
    ~
    i = 1, \ldots, N    
\end{equation}
where $\mu_j < \mu_i$, for $i < j$, $\mu_\text{max}$ confines the system near the physical subspace, and $\mu_\text{min}$ permits the greatest extension.

For a harmonic penalty $\tfrac12\mu\,\mathcal{L}$ with $\mathcal{L}=\sum_\delta x_\delta^2$ at the separable Hamiltonian, equipartition~\cite{tuckerman2010statmech} gives
\begin{align}
    &\langle x_{i,\delta}^2 \rangle = k_B T/\mu,
    \qquad    
    \langle\mathcal{L}\rangle = N_\text{dof}\,k_BT/\mu, 
    \label{eq:equipartition}\\
    &\mathrm{Var}(\mathcal{L}) = 2N_\text{dof}\,(k_BT/\mu)^2,
    \label{eq:variance}
\end{align}
so that the root-mean-square displacement $\langle \overline{\bf x}_+^2\rangle^{1/2}\propto\mu^{-1/2}$ and the mean penalty energy $\tfrac12 N_\text{dof}\,k_BT$ is $\mu$-independent. Acceptance is controlled by fluctuations of $\mathcal{L}$, following the Gaussian-energy analysis of Ref.~\cite{kofke2002acceptance}.

Recall that for adjacent replicas for the non-separable Hamiltonian, the exchange exponent $\Delta_{i,i+1}$ is represented as 
\begin{equation}
\label{eq:delta}
    \Delta_{i,i+1} = \frac{\mu_i - \mu_{i+1}}{2k_BT}\left(\mathcal{L}(\overline{\bf x}_{+,i}) - \mathcal{L}(\overline{\bf x}_{+,i+1})\right), 
\end{equation}
as in Eq.~\eqref{eq:acceptance_criterion}. 
When $\mu_{i}$ is sufficiently large, Eq.~\eqref{eq:delta} is asymptotically equivalent to Eq.~\eqref{eq:acceptance_criterion_separable}, the acceptance criterion for the separable case. This identification enables to apply the expectations from Eq.~\eqref{eq:equipartition} to Eq.~\eqref{eq:delta}, yielding
\begin{equation}
\label{eq:delta-mean}
    \langle\Delta_{i,i+1}\rangle
    =
    -\frac{N_\text{dof}}{2}\,\frac{(\rho-1)^2}{\rho}
\end{equation}
with $\rho=\mu_i/\mu_{i+1}>1$.
If the two replicas are independent, the corresponding standard deviation is
\begin{equation}
\label{eq:delta-std}
    \sigma_{\Delta}
    =
    \sqrt{\frac{N_\text{dof}}{2}}\,\frac{\rho-1}{\rho}\,\sqrt{1+\rho^2}.
\end{equation}
Both $\langle\Delta_{i,i+1}\rangle$ and $\sigma_{\Delta}$ depend on $\rho$ and $N_\text{dof}$ alone, and are independent of the absolute scale of 
$\mu$~\cite{kofke2002acceptance}. 
For closely spaced replicas ($\rho=1+\varepsilon$, $\varepsilon\ll 1$), one has $|\langle\Delta_{i,i+1}\rangle|\sim\tfrac12 N_\text{dof}\,\varepsilon^2$ and $\sigma_{\Delta}\sim\sqrt{N_\text{dof}}\,\varepsilon$; geometric spacing designed for $|\langle\Delta_{i,i+1}\rangle|\sim O(1)$ therefore places mean and fluctuation on the same scale.

We emphasize that for non-separable replicas, especially with small $\mu$, these expressions are heuristics rather than exact results; the approximately balanced acceptance rates in Fig.~\ref{fig:ala2_rex_exchange}(a) provide empirical support for the geometric schedule in the present system.

\end{document}